\documentclass[12pt,cite,a4paper]{article}
\usepackage{amsmath}
\usepackage{myart}
\usepackage{graphicx}
\usepackage{amsfonts}
\usepackage{amssymb}

\begin{document}
\hfill\hbox{PM/03-34}

\hfill\hbox{December 2003}

\bigskip

\begin{center}
{\Large \textbf{Higher Equations of Motion in Liouville Field Theory}}

\vspace{0.5cm}

{\large Contribution to the proceedings of the VI International Conference
``CFT and Integrable Models'', Chernogolovka, Russia, September 2002.}

\vspace{1.0cm}

{\large Al.Zamolodchikov}\footnote{On leave of absence from Institute of
Theoretical and Experimental Physics, B.Cheremushkinskaya 25, 117259 Moscow, Russia.}

\vspace{0.2cm}

Laboratoire de Physique Math\'ematique\footnote{Laboratoire Associ\'e au CNRS URA-768}

Universit\'e Montpellier II

Pl.E.Bataillon, 34095 Montpellier, France
\end{center}

\vspace{1.0cm}

\textbf{Abstract}

An infinite set of operator-valued relations in Liouville field theory is
established. These relations are enumerated by a pair of positive integers
$(m,n)$, the first $(1,1)$ representative being the usual Liouville equation
of motion. The relations are proven in the framework of conformal field theory
on the basis of exact structure constants in the Liouville operator product
expansions. Possible applications in 2D gravity are discussed.

\section{Introduction}

To give an idea about the subject of this lecture, let's start with a simple
observation. Consider the classical 2D Liouville equation
\begin{equation}
\partial\bar\partial\varphi=Me^{\varphi}\label{cLiouv}%
\end{equation}
Here as usual the complex coordinates $z=x+iy$ and $\bar z=x-iy$ are implied,
so that $\partial=\left(  \partial_{x}-i\partial_{y}\right)  /2$ and
$\bar\partial=\left(  \partial_{x}+i\partial_{y}\right)  /2$. An irrelevant
parameter $M$ is introduced here to make equations below more transparent. It
follows directly from (\ref{cLiouv}) that the ``classical stress tensor''
components
\begin{align}
T^{\text{(c)}}  &  =-\frac14(\partial\varphi)^{2}+\frac12\partial^{2}%
\varphi\label{cstress}\\
\bar T^{\text{(c)}}  &  =-\frac14(\bar\partial\varphi)^{2}+\frac12\bar
\partial^{2}\varphi\nonumber
\end{align}
are holomorphic and antiholomorphic functions of $z$ respectively, i.e.,
$\bar\partial T^{\text{(c)}}=0$; $\partial\bar T^{\text{(c)}}=0$.

Consider the infinite series of fields $V_{n}^{\left(  \text{c}\right)
}=e^{\left(  1-n\right)  \varphi/2}$, $n=1,2,3,\ldots$. First representatives
of this series $1,e^{-\varphi/2},e^{-\varphi},e^{-3\varphi/2},e^{-2\varphi
},\ldots$. are straightforwardly verified to satisfy the following
differential equations
\begin{align}
\partial\cdot1 &  =0\nonumber\\
\left(  \partial^{2}+T^{\text{(c)}}\right)  e^{-\varphi/2} &  =0\nonumber\\
\left(  \partial^{3}+4T^{\text{(c)}}\partial+2\partial T^{\text{(c)}}\right)
e^{-\varphi} &  =0\nonumber\\
\left(  \partial^{4}+10T^{\text{(c)}}\partial^{2}+10\partial T^{\text{(c)}%
}\partial+\left(  9T^{\text{(c)}2}+3\partial^{2}T^{\text{(c)}}\right)
\right)  e^{-3\varphi/2} &  =0\nonumber\\
\left(  \partial^{5}+20T^{\text{(c)}}\partial^{3}+30\partial T^{\text{(c)}%
}\partial^{2}+\left(  64T^{\text{(c)}2}+18\partial^{2}T^{\text{(c)}}\right)
\partial+\left(  64T^{\text{(c)}}\partial T^{\text{(c)}}\right)
+4\partial^{3}T^{\text{(c)}}\right)  e^{-2\varphi} &  =0\nonumber\\
&  \ldots\label{cnull}%
\end{align}
and the same equations with $\partial$ and $T^{\text{(c)}}$ replaced by
$\bar\partial$ and $\bar T^{\text{(c)}}$. It is commonly believed that this
series of relations goes on ad inf and for each $V_{n}^{\left(  \text{c}%
\right)  }$, $n=0,1,2,\ldots$ a unique differential operator exists (albeit
its form is unknown in general)
\begin{equation}
D_{n}^{\text{(c)}}=\partial^{n}+\ldots+(n-1)\partial^{n-2}T^{\text{(c)}%
}\label{Dc}%
\end{equation}
(the coefficients being graded polynomials in $T^{\text{(c)}}$ and its
derivatives) such that
\begin{equation}
D_{n}^{\text{(c)}}V_{n}^{\left(  \text{c}\right)  }=\bar D_{n}^{\text{(c)}%
}V_{n}^{\left(  \text{c}\right)  }=0\label{cnullv}%
\end{equation}
The ``left'' operator $\bar D_{n}^{\text{(c)}}=\bar\partial^{n}+\ldots
+(n-1)\partial^{n-2}\bar T^{\text{(c)}}$ differs from (\ref{Dc}) in the same
replacement $\partial\rightarrow\bar\partial$ and $T^{\text{(c)}}%
\rightarrow\bar T^{\text{(c)}}$.

Let us now take the fields $\varphi V_{n}^{\text{(c)}}$, $n=1,2,3,\ldots$,
i.e., $\varphi$, $\varphi e^{-\varphi/2}$, $\varphi e^{-\varphi}$, $\varphi
e^{-3\varphi/2}$, $\varphi e^{-2\varphi}$, etc.. Then,
\begin{align}
\bar D_{1}^{\text{(c)}}D_{1}^{\text{(c)}}\varphi &  =Me^{\varphi}\nonumber\\
\bar D_{2}^{\text{(c)}}D_{2}^{\text{(c)}}\left(  \varphi e^{-\varphi
/2}\right)   &  =-M^{2}e^{3\varphi/2}\nonumber\\
\bar D_{3}^{\text{(c)}}D_{3}^{\text{(c)}}\left(  \varphi e^{-\varphi}\right)
&  =3M^{3}e^{2\varphi}\label{chigher}\\
\bar D_{4}^{\text{(c)}}D_{4}^{\text{(c)}}\left(  \varphi e^{-3\varphi
/2}\right)   &  =-18M^{4}e^{5\varphi/2}\nonumber\\
\bar D_{5}^{\text{(c)}}D_{5}^{\text{(c)}}\left(  \varphi e^{-2\varphi
}\right)   &  =180M^{5}e^{3\varphi}\nonumber
\end{align}
The first line is just the Liouville equation of motion. The next are verified
by a straightforward differential calculus with the use of the Liouville
equation. It seems natural to surmise in general
\begin{equation}
\bar D_{n}^{\text{(c)}}D_{n}^{\text{(c)}}\left(  \varphi e^{(1-n)\varphi
/2}\right)  =B_{n}^{\text{(c)}}e^{(1+n)\varphi/2}\label{cnhigh}%
\end{equation}
with
\begin{equation}
B_{n}^{\text{(c)}}=2(-)^{n+1}n!(n-1)!\left(  \frac M2\right)  ^{n}\label{Bc}%
\end{equation}

It is the matter of this lecture to demonstrate that similar, but somewhat
extended set of relations holds between the operators in quantum Liouville
field theory \cite{Teschner}. The talk is organized as follows. In the next
section we describe very briefly the Liouville field theory, mainly to
introduce the notations and remind the basic features. In sect. 3 the
degenerate primary fields $V_{m,n}$ in Liouville field theory are discussed in
some more detail and the basic operators $D_{m,n}$ are definded. Sect.4
introduces the corresponding logarithmic Liouville primaries $V_{m,n}^{\prime
}$, and lines out some of their basic properties. These fields are basically
the derivatives of general Liouville exponential $V_{\alpha}$ at the point
$\alpha_{m,n}$. Such fields were considered in the context of 2D Liouville
gravity in paper \cite{Polchinski}. The main proposition about the effect of
$D_{m,n}\bar D_{m,n}$ on $V_{m,n}^{\prime}$ is proven. Then, in sect.5 the
quantum higher equations of motion are carried out. Some applications are
finally discussed in a pure algebraic problem (sections 6 and 7) as well as in
the minimal Liouville gravity (sect.8)

\section{Liouville field theory}

Liouville field theory (LFT) is defined by the Lagrangian density \cite{Teschner}%

\begin{equation}
\mathcal{L}=\frac1{4\pi}\left(  \partial_{a}\phi\right)  ^{2}+\mu e^{2b\phi
}\label{LFT}%
\end{equation}
where $\phi$ is the Liouville field and $b$ is the dimensionless coupling.
Scale parameter $\mu$ is called the cosmological constant. LFT is a particular
and rather specific example of conformal field theory (CFT). Conformal
invariance follows from the existence of holomorphic and antiholomorphic
components of the stress tensor
\begin{align}
T(z) &  =-\left(  \partial\phi\right)  ^{2}+Q\partial^{2}\phi\label{T}\\
\bar T(\bar z) &  =-\left(  \bar\partial\phi\right)  ^{2}+Q\bar\partial
^{2}\phi\nonumber
\end{align}
where the convenient combination
\begin{equation}
Q=b^{-1}+b\label{Q}%
\end{equation}
($Q$ is called traditionally the background charge) is introduced. This stress
tensor gives rise to the right and left Virasoro symmetries with central
charge
\begin{equation}
c_{L}=1+6Q^{2}\label{cL}%
\end{equation}
in the space of states and in the space of local fields.

Action (\ref{LFT}) describes the local properties of the system. Different
problems in LFT are specified by the boundary conditions and correspond to
different geometries. E.g., in the so called spherical geometry the field
$\phi$ is defined over all two-dimensional plane and has the following
behavior at $\left|  x\right|  \rightarrow\infty$%
\begin{equation}
\phi(x)\sim-2Q\log\left|  x\right|  +O(1)\;\;\;\;\;\;\;\;\left|  x\right|
\rightarrow\infty\label{sphere}%
\end{equation}

Formally Lagrangian (\ref{LFT}) leads to the following equation of motion for
the field $\phi$
\begin{equation}
\partial\bar\partial\phi=\pi b\mu e^{2b\phi}\label{eq11}%
\end{equation}
Classical Liouville equation (\ref{cLiouv}) is recovered in the classical
limit $b^{2}\rightarrow0$ of the quantum construction. In this limit the
classical fields and parameters are related to the ones of this section as
\begin{align}
2b\phi &  \rightarrow\varphi\nonumber\\
T  &  \rightarrow b^{-2}T^{\text{(c)}}\label{climit}\\
2\pi\mu b^{2}  &  \rightarrow M\nonumber
\end{align}

In practice it is hard to construct the solution of LFT starting directly from
equation of motion (\ref{eq11}). It turns out more productive to begin with
certain set of rather natural premises which permit to construct LFT as a
complete and coherent set of CFT-consistent correlation functions of basic
primary fields. Then it can be demonstrated that properly defined field $\phi$
(which is not a primary field in the CFT context) satisfies eq.(\ref{eq11})
\cite{DO}. This observation makes it natural to interpret the whole
construction as the exact solution of LFT.

First, one presumes the existence of the continuous set of basic local primary
fields $V_{\alpha}$ parameterized by (in general complex) parameter $\alpha$.
This parameter is usually supposed to satisfy the restriction
\begin{equation}
\operatorname*{Re}\alpha<Q/2\label{Seiberg}%
\end{equation}
called the Seiberg bound \cite{Seiberg}. Operators $V_{\alpha}$ are
interpreted as the properly regularized exponentials in the Liouville field
$\phi$
\begin{equation}
V_{\alpha}(x)=e^{2\alpha\phi(x)}\label{Va}%
\end{equation}
With respect to the conformal symmetry these fields are scalar primaries of
dimensions $\left(  \Delta_{\alpha},\Delta_{\alpha}\right)  $
\begin{equation}
\Delta_{\alpha}=\alpha(Q-\alpha)\label{Da}%
\end{equation}

It is convenient to normalize these fields through the two-point function
$\left\langle V_{\alpha}(x)V_{\alpha}(0)\right\rangle =S(\alpha)\left(  x\bar
x\right)  ^{-2\Delta_{a}}$ with \cite{zz1}
\begin{equation}
S(\alpha)=\left(  \pi\mu\gamma(b^{2})\right)  ^{(Q-2\alpha)/b}\frac
{\gamma(2\alpha b-b^{2})}{b^{2}\gamma(2-2\alpha b^{-1}+b^{-2})}\label{D}%
\end{equation}
Here and below the standard notation $\gamma(x)=\Gamma(x)/\Gamma(1-x)$ is
used. In this normalization the operator product expansion algebra of these
basic primaries is determined by the three-point function in spherical
geometry
\begin{equation}
\left\langle V_{\alpha_{1}}(x_{1})V_{\alpha_{2}}(x_{2})V_{\alpha_{3}}%
(x_{3})\right\rangle =\frac{C(\alpha_{1},\alpha_{2},\alpha_{3})}{(x_{12}\bar
x_{12})^{\Delta_{1}+\Delta_{2}-\Delta_{3}}(x_{23}\bar x_{23})^{\Delta
_{3}+\Delta_{2}-\Delta_{1}}(x_{31}\bar x_{31})^{\Delta_{3}+\Delta_{1}%
-\Delta_{2}}}\label{V3}%
\end{equation}
where the structure constant $C(\alpha_{1},\alpha_{2},\alpha_{3})$ has the
following explicit form \cite{DO, Teschner}
\begin{align}
C(\alpha_{1},\alpha_{2},\alpha_{3}) &  =\left(  \pi\mu\gamma(b^{2}%
)b^{2-2b^{2}}\right)  ^{(Q-\sum_{i=1}^{3}\alpha_{i})/b}\times\nonumber\\
&  \ \frac{\Upsilon^{\prime}(0)\Upsilon(2\alpha_{1})\Upsilon(2\alpha
_{2})\Upsilon(2\alpha_{2})}{\Upsilon(\sum_{i=1}^{3}\alpha_{i}-Q)\Upsilon
(\alpha_{1}+\alpha_{2}-\alpha_{3})\Upsilon(\alpha_{2}+\alpha_{3}-\alpha
_{1})\Upsilon(\alpha_{3}+\alpha_{1}-\alpha_{2})}\label{G3}%
\end{align}
Special function $\Upsilon(x)$, introduced in \cite{zz1}, is closely related
to the Barnes double gamma function \cite{Barnes}. It satisfies the following
defining functional relations
\begin{align}
\Upsilon(x+b) &  =\gamma(bx)b^{1-2bx}\Upsilon(x)\label{Upsilon}\\
\Upsilon(x+b^{-1}) &  =\gamma(b^{-1}x)b^{2b^{-1}x-1}\Upsilon(x)\nonumber
\end{align}

It is important to note that these basic characteristics are analytic
(meromorphic) functions of parameters $\alpha$. In particular, from the
analytic point of view nothing special happens at the Seiberg bound
(\ref{Seiberg}). This means that the latter is more an interpretational
concept then a genuine analytic bound. In fact it is possible to define
formally the ``exponential'' operators (\ref{Va}) with ${\operatorname*{Re}%
\alpha}\geq Q/2$ as the analytic continuation from the ``physical'' region
$\operatorname*{Re}\alpha<Q/2$. Under this continuation the reflection
relation between the exponential operators
\begin{equation}
V_{\alpha}=S(\alpha)V_{Q-\alpha}\label{refl}%
\end{equation}
holds \cite{Teschner, zz1}.

\section{Degenerate Liouville primaries}

Among the exponential primaries $V_{\alpha}$ there is a discrete subset
$V_{m,n}=V_{\alpha_{m,n}}$ where $\left(  m,n\right)  $ is a pair of positive
integers. They correspond to special values of the parameter $\alpha
=\alpha_{m,n}$
\begin{equation}
\alpha_{m,n}=-(m-1)b^{-1}/2-(n-1)b/2\label{amn}%
\end{equation}
These fields are called degenerate, because the corresponding dimensions
\begin{equation}
\Delta_{m,n}=\frac{Q^{2}}4-\frac{\left(  mb^{-1}+nb\right)  ^{2}}4\label{dmn}%
\end{equation}
are the Kac dimensions \cite{Kac} of the degenerate representations of the
Virasoro algebra with central charge (\ref{cL}). The corresponding (say
``right'') null-vectors appear at level $mn$. They can be represented as the
action of certain operators $D_{m,n}$ on the highest weight vector $V_{m,n}$.
Operator $D_{m,n}$ is a graded polynomial of level $mn$ in the right Virasoro
algebra generators $L_{n}$%
\begin{equation}
D_{m,n}=L_{-1}^{mn}+d_{1}^{m,n}(b)L_{-2}L_{-1}^{mn-2}+\ldots\label{Dmn}%
\end{equation}
Of course there are also ``left'' null vectors created by the left operators
$\bar D_{m,n}$ (with all ``right'' Virasoro generators $L_{n}$ replaced by the
``left'' ones $\bar L_{n}$).

Up to now I ignore the generic form of the coefficients $d_{i}^{m,n}(b)$ in
(\ref{Dmn}). Level by level they can be carried out by a straightforward
algebra \cite{BPZ}. E.g., at level 2 one finds (below we don't quote the
corresponding ``dual'' operators since $D_{m,n}$ is related to $D_{n,m}$
simply by the substitution $b\rightarrow b^{-1}$)
\begin{equation}
D_{1,2}=L_{-1}^{2}+b^{2}L_{-2}\label{D2}%
\end{equation}
At level 3
\begin{equation}
D_{1,3}=L_{-1}^{3}+4b^{2}L_{-2}L_{-1}+2b^{2}\left(  1+2b^{2}\right)
L_{-3}\label{D3}%
\end{equation}
and at fourth
\begin{align}
D_{1,4} &  =L_{-1}^{4}+10b^{2}L_{-2}L_{-1}^{2}+2b^{2}\left(  5+12b^{2}\right)
L_{-3}L_{-1}+9b^{4}L_{-2}^{2}+6b^{2}\left(  1+4b^{2}+6b^{4}\right)
L_{-4}\nonumber\\
D_{2,2} &  =L_{-1}^{4}+2\left(  b^{-2}+b^{2}\right)  L_{-2}L_{-1}^{2}+2\left(
b^{-2}+3+b^{2}\right)  L_{-3}L_{-1}+\left(  b^{-4}-2+b^{4}\right)  L_{-2}%
^{2}+\label{D4}\\
&  +3\left(  b^{-2}+2+b^{2}\right)  L_{-4}\nonumber
\end{align}

One of the basic assumptions in the LFT construction \cite{GervaisNeveu},
which allows in fact to reconstruct the whole structure including operator
product algebra (\ref{G3}) \cite{TT}, is that for each $(m,n)$ the
null-vectors created by the action of $D_{m,n}$ or $\bar D_{m,n}$ vanish
\begin{equation}
D_{m,n}V_{m,n}=\bar D_{m,n}V_{m,n}=0\label{null}%
\end{equation}
With standard interpretation of the Virasoro generators in terms of the stress
tensor, $L_{-1}\rightarrow\partial$ and $(n-2)!L_{-n}\rightarrow\partial
^{n-2}T$ at $n>1$, these equations can be viewed as linear differential
equations for $V_{m,n}$, the coefficients being graded polynomials in $T$ and
its derivatives. Only the subset $(1,n)$ allows a smooth classical limit. It
reduces to (\ref{cnull}) at $b\rightarrow0$, relation (\ref{climit}) between
the quantum and classical stress tensor components being taken into account.

\section{Logarithmic degenerate fields}

Define the logarithmic field
\begin{equation}
V_{\alpha}^{\prime}=\frac12\frac\partial{\partial\alpha}V_{\alpha}=\phi
e^{2\alpha\phi}\label{Vprim}%
\end{equation}
Corresponding to the degenerate set $V_{m,n}$ we have the following discrete
set of logarithmic fields
\begin{equation}
V_{m,n}^{\prime}=V_{\alpha}^{\prime}|_{\alpha=\alpha_{m,n}}\label{Vmnprim}%
\end{equation}
For example, $V_{1,1}^{\prime}$ is the Liouville field $\phi$ itself,
$V_{1,2}^{\prime}=\phi e^{-b\phi}$ etc.

\textbf{Proposition:}%

\begin{equation}
D_{m,n}\bar D_{m,n}V_{m,n}^{\prime}\label{prim}%
\end{equation}
is a primary field.

\textbf{Proof: }Consider first $\bar D_{m,n}V_{\alpha}$ in the vicinity of
$\alpha=\alpha_{m,n}$. From analyticity in $\alpha$, mentioned above, we have
$\bar D_{m,n}V_{\alpha}=\left(  \alpha-\alpha_{m,n}\right)  A_{m,n}+O\left(
(\alpha-\alpha_{m,n})^{2}\right)  $ where $A_{m,n}$ is an operator of
dimension $(\Delta_{m,n}+mn,\Delta_{m,n})$ which is no more left primary but
still a right primary. Hence $D_{m,n}A_{m,n}=2D_{m,n}\bar D_{m,n}%
V_{m,n}^{\prime}$ is also a right primary. In the above consideration one
could inverse the roles of $D_{m,n}$ and $\bar D_{m,n}$ and therefore
$D_{m,n}\bar D_{m,n}V_{m,n}^{\prime}$ is also a left primary$\Box$

Thus $D_{m,n}\bar D_{m,n}V_{m,n}^{\prime}$ is a primary of dimension
$(\tilde\Delta_{m,n},\tilde\Delta_{m,n})$ where $\tilde\Delta_{m,n}%
=\Delta_{m,n}+mn$. In the primary field spectrum of the Liouville field theory
there is an exponential field $V_{\alpha}$ with $\alpha=\tilde\alpha_{m,n}$
\begin{equation}
\tilde\alpha_{m,n}=-(m-1)/2+(n+1)b/2\label{alphatilde}%
\end{equation}
which we denote $\tilde V_{m,n}=V_{\tilde\alpha_{m,n}}$, of the same dimension
$(\tilde\Delta_{m,n},\tilde\Delta_{m,n})$. The main goal of this work is to
establish the operator-valued relation
\begin{equation}
D_{m,n}\bar D_{m,n}V_{m,n}^{\prime}=B_{m,n}\tilde V_{m,n}\label{heq}%
\end{equation}
and calculate explicitly the numerical coefficients $B_{m,n}$.

\section{Higher equations of motion}

In what follows it will prove convenient to replace sometimes parameters
$\alpha$ in the fields $V_{\alpha}$ by slightly different ones $\lambda$
defined as
\begin{equation}
\alpha=Q/2-\lambda\label{lambda}%
\end{equation}
The reflection $\alpha\rightarrow Q-\alpha$ is simply $\lambda\rightarrow
-\lambda$, dimensions $\Delta_{\alpha}=Q^{2}/4-\lambda^{2}$ being even
functions of $\lambda$. Introducing the notation
\begin{equation}
\lambda_{m,n}=(mb^{-1}+nb)/2\label{lmn}%
\end{equation}
let's define also the following polynomials for every pair $(m,n)$ of positive
integers
\begin{equation}
p_{m,n}(x)=\prod_{r,s}(x-\lambda_{r,s})\label{pmn}%
\end{equation}
where the pair of integers $(r,s)$ runs over the set
\begin{align}
r  &  =-m+1,-m+3,\ldots,m-2,m-1\label{set}\\
s  &  =-n+1,-n+3,\ldots,n-2,n-1\nonumber
\end{align}
Polynomial $p_{m,n}(x)$ has the order $mn$ in $x$ and the same parity as the
product $mn$.

Operator-valued relation (\ref{heq}) means that for every multipoint
correlation function the equality
\begin{equation}
\left\langle D_{m,n}\bar D_{m,n}V_{m,n}^{\prime}V_{\alpha_{1}}\ldots
V_{\alpha_{N}}\right\rangle =B_{m,n}\left\langle \tilde V_{m,n}V_{\alpha_{1}%
}\ldots V_{\alpha_{N}}\right\rangle \label{hem}%
\end{equation}
holds. Due to the conformal invariance of the theory it is enough to verify
this relation for the structure constant in the operator product expansions,
or equivalently, for the three-point function. Hence, we have to compare
$C_{m,n}^{\prime}=\left\langle D_{m,n}\bar D_{m,n}V_{m,n}^{\prime}%
(x)V_{\alpha_{1}}(x_{1})V_{\alpha_{2}}(x_{2})\right\rangle $ and $\tilde
C_{m,n}=\left\langle \tilde V_{m,n}(x)V_{\alpha_{1}}(x_{1})V_{\alpha_{2}%
}(x_{2})\right\rangle $. While the second three point function is obtained
from (\ref{V3}) straightforwardly
\begin{equation}
\tilde C_{m,n}=\frac{C(\tilde\alpha_{m,n},\alpha_{1},\alpha_{2})}{\left|
x_{12}\right|  ^{2\Delta_{1}+2\Delta_{2}-2\tilde\Delta_{m,n}}\left|
x-x_{1}\right|  ^{2\tilde\Delta_{m,n}+2\Delta_{1}-2\Delta_{2}}\left|
x-x_{2}\right|  ^{2\tilde\Delta_{m,n}+2\Delta_{2}-2\Delta_{1}}}\label{Ctilde}%
\end{equation}
evaluation of the first one involves some subtleties. First notice that
$C(\alpha,\alpha_{1},\alpha_{2})$ has a first order zero as $\alpha
\rightarrow\alpha_{m,n}$. Thus
\begin{align}
\  &  \left\langle V_{m,n}^{\prime}(x)V_{\alpha_{1}}(x_{1})V_{\alpha_{2}%
}(x_{2})\right\rangle =\label{long}\\
&  \ \frac{\partial C(\alpha,\alpha_{1},\alpha_{2})/\partial\alpha
|_{\alpha=\alpha_{m,n}}}{2\left|  x_{12}\right|  ^{2\Delta_{1}+2\Delta
_{2}-2\Delta_{m,n}}\left|  x-x_{1}\right|  ^{2\Delta_{m,n}+2\Delta_{1}%
-2\Delta_{2}}\left|  x-x_{2}\right|  ^{2\Delta_{m,n}+2\Delta_{2}-2\Delta_{1}}%
}\nonumber
\end{align}
Now, the action of $D_{m,n}$ and $\bar D_{m,n}$ affect only the coordinate
dependence of the correlation function, acting on the holomorphic and
antiholomophic factors independently. E.g., $D_{m,n}$ reduces to a certain
differential operator
\begin{equation}
\mathcal{P}_{m,n}=(\partial/\partial x)^{mn}+\ldots\label{ddmn}%
\end{equation}
in $x$, $x_{1}$ and $x_{2}$ of maximal order $mn$ (the term written up in
(\ref{ddmn}) explicitly), the coefficients being rational functions in the
coordinates. Kinematics of conformal field theory ensures that (remember that
$D_{m,n}$ produces a primary state of dimension $\tilde\Delta_{m,n}$ after
acting on a primary of dimension $\Delta_{m,n}$)
\begin{align}
&  \ \ \mathcal{P}_{m,n}(x_{12})^{\Delta_{m,n}-\Delta_{1}-\Delta_{2}}%
(x-x_{1})^{\Delta_{2}-\Delta_{1}-\Delta_{m,n}}(x-x_{2})^{\Delta_{1}-\Delta
_{2}-\Delta_{m,n}}\nonumber\\
\  &  =P_{m,n}(\Delta_{1},\Delta_{2})(x_{12})^{\tilde\Delta_{m,n}-\Delta
_{1}-\Delta_{2}}(x-x_{1})^{\Delta_{2}-\Delta_{1}-\tilde\Delta_{m,n}}%
(x-x_{2})^{\Delta_{1}-\Delta_{2}-\tilde\Delta_{m,n}}\label{Pmn}%
\end{align}
where $P_{m,n}$ is a polynomial in $\Delta_{1}$ and $\Delta_{2}$. Again, it is
easy to argue that (consider that each appearance of $L_{-n}$ with $n>1$ in
the explicit expression of $D_{m,n}$ gives at most a first power of $\Delta$
while decreasing the order of $\partial$ in the differential operator
$\ \mathcal{P}_{m,n}$ by at least two) the maximal order (in $\Delta_{1}$ and
$\Delta_{2}$) term in $P_{m,n}(\Delta_{1},\Delta_{2})$ is provided by the term
$(\partial/\partial x)^{mn}$ explicitly written down in (\ref{ddmn}). Thus
\begin{equation}
P_{m,n}(\Delta_{1},\Delta_{2})=(\Delta_{2}-\Delta_{1})^{mn}+\text{lower order
terms}\label{lorder}%
\end{equation}
Moreover, it is well known since \cite{BPZ}, that the left hand side of
eq.(\ref{Pmn}) kinematically vanishes if the dimensions $\Delta_{1}$ and
$\Delta_{2}$ satisfy the fusion rules of the degenerate operator $V_{m,n}$.
Let me remind that it is this fact that allows to interpret the fusion rules
as a consequence of the vanishing of the null-vector (\ref{null}). In terms of
$\lambda_{1}$ and $\lambda_{2}$ the fusion relations are conveniently
summarized as \cite{BPZ}
\begin{align}
\lambda_{+} &  =\lambda_{1}+\lambda_{2}=\lambda_{r,s}\;\;\text{or}%
\label{lambafusion}\\
\lambda_{-} &  =\lambda_{2}-\lambda_{1}=\lambda_{r.s}\nonumber
\end{align}
with any pair $(r,s)$ from the set (\ref{set}). Therefore the polynomial
$P_{m,n}(\Delta_{1},\Delta_{2})$ is proportional to $p_{m,n}(\lambda
_{+})p_{m,n}(\lambda_{-})$. Comparing the maximal order term in the product
(\ref{pmn}) with (\ref{lorder}) we see that in fact it is equal to this
product
\begin{equation}
P_{m,n}(\Delta_{1},\Delta_{2})=p_{m,n}(\lambda_{1}+\lambda_{2})p_{m,n}%
(\lambda_{2}-\lambda_{1})\label{Ppp}%
\end{equation}

The same arguments are applied to the action of $\bar D_{m,n}.$ Thus
\begin{equation}
C_{m,n}^{\prime}=\frac{P_{m,n}^{2}(\Delta_{1},\Delta_{2})\partial
C(\alpha,\alpha_{1},\alpha_{2})/\partial\alpha|_{\alpha=\alpha_{m,n}}%
}{2\left|  x_{12}\right|  ^{2\Delta_{1}+2\Delta_{2}-2\tilde\Delta_{m,n}%
}\left|  x-x_{1}\right|  ^{2\tilde\Delta_{m,n}+2\Delta_{1}-2\Delta_{2}}\left|
x-x_{2}\right|  ^{2\tilde\Delta_{m,n}+2\Delta_{2}-2\Delta_{1}}}\label{Cmnprim}%
\end{equation}
With explicit expression (\ref{G3}) for the structure constant this gives
(recall that $\alpha_{m,n}=Q/2-\lambda_{m,n}$ and $\tilde\alpha_{m,n}%
=Q/2-\lambda_{m,-n}$)
\begin{align}
\  &  \frac{C_{m,n}^{\prime}}{\tilde C_{m,n}P_{m,n}^{2}(\Delta_{1},\Delta
_{2})}=\frac{\partial C(\alpha,\alpha_{1},\alpha_{2})/\partial\alpha
|_{\alpha=\alpha_{m,n}}}{2C(\tilde\alpha_{m,n},\alpha_{1},\alpha_{2})}=\left(
\pi\mu\gamma(b^{2})b^{2-2b^{2}}\right)  ^{n}\frac{\Upsilon^{\prime}%
(2\alpha_{m,n})}{\Upsilon(2\tilde\alpha_{m,n})}\times\label{r}\\
&  \ \ \frac{\Upsilon(Q/2-\lambda_{+}-\tilde\lambda_{m,n})\Upsilon
(Q/2+\lambda_{+}-\tilde\lambda_{m,n})\Upsilon(Q/2-\lambda_{-}-\tilde
\lambda_{m,n})\Upsilon(Q/2+\lambda_{-}-\tilde\lambda_{m,n})}{\Upsilon
(Q/2-\lambda_{+}-\lambda_{m,n})\Upsilon(Q/2+\lambda_{+}-\lambda_{m,n}%
)\Upsilon(Q/2-\lambda_{-}-\lambda_{m,n})\Upsilon(Q/2+\lambda_{-}-\lambda
_{m,n})}\nonumber
\end{align}
where we've introduced $\tilde\lambda_{m,n}=\lambda_{m,-n}$.

The shift relations (\ref{Upsilon}) allow to establish the following identity
for the $\Upsilon$-functions%

\begin{equation}
\frac{\Upsilon(Q/2+x-\lambda_{m,-n})\Upsilon(Q/2-x-\lambda_{m,-n})}%
{\Upsilon(Q/2+x-\lambda_{m,n})\Upsilon(Q/2-x-\lambda_{m,n})}=\frac{(-1)^{mn}%
}{p_{m,n}^{2}(x)}\label{Urelation}%
\end{equation}
Thus, all the dependence on the parameters $\alpha_{1}$ and $\alpha_{2}$
disappears in the ratio of the structure constants leaving a constant, which
is interpreted as the constant $B_{m,n}$ in the operator relation (\ref{heq})
\begin{equation}
\frac{\left\langle D_{m,n}\bar D_{m,n}V_{m,n}^{\prime}(x)V_{\alpha_{1}}%
(x_{1})V_{\alpha_{2}}(x_{2})\right\rangle }{\left\langle \tilde V_{m,n}%
(x)V_{\alpha_{1}}(x_{1})V_{\alpha_{2}}(x_{2})\right\rangle }=B_{m,n}=\left(
\pi\mu\gamma(b^{2})b^{2-2b^{2}}\right)  ^{n}\frac{\Upsilon^{\prime}%
(2\alpha_{m,n})}{\Upsilon(2\tilde\alpha_{m,n})}\label{Bmn}%
\end{equation}
The ratio in the right hand side can be further simplified to
\begin{equation}
B_{m,n}=\left(  \pi\mu\gamma(b^{2})\right)  ^{n}b^{1+2n-2m}\gamma
(m-nb^{2})\prod_{\substack{k=1-n \\l=1-m \\(k,l)\neq(0,0) }}^{\substack{m-1
\\n-1 }}(lb^{-1}+kb)\label{klprod}%
\end{equation}

Let's consider few properties of this expression.

\textbf{1. Classical limit.} Only the $B_{1,n}$ series of equations admits the
classical limit $b\rightarrow0$. In this case
\begin{equation}
B_{1,n}=\left(  \pi\mu\gamma(b^{2})\right)  ^{n}\frac{b^{4n-3}(-)^{n-1}%
((n-1)!)^{2}}{\gamma(nb^{2})}\label{B1n}%
\end{equation}
and in the limit
\begin{equation}
B_{1,n}\rightarrow_{b\rightarrow0}\left(  \pi\mu\gamma b^{2}\right)  ^{n}%
\frac{(-)^{n-1}n!(n-1)!}b\;\label{B1nlimit}%
\end{equation}
This conforms expression (\ref{Bc}) (remember that $\varphi\sim2b\phi$ and
$\pi\mu\gamma(b^{2})\rightarrow M$ at $b\rightarrow0$).

\textbf{2. Duality.} Replace the field $\tilde V_{m,n}=V_{m,-n}$ with the
``reflected'' one $S(\tilde\alpha_{m,n})V_{Q-\tilde\alpha_{m,n}}$ from
eq.(\ref{refl}). Explicitly
\begin{equation}
S(\tilde\alpha_{m,n})=\left(  \pi\mu\gamma(b^{2})\right)  ^{mb^{-2}-n}%
\frac{\gamma(1-m+nb^{2})}{b^{2}\gamma(1+mb^{-2}-n)}\label{Dalphamn}%
\end{equation}
so that
\begin{equation}
B_{n,m}S(\tilde\alpha_{n,m})=\left(  \pi\tilde\mu\gamma(b^{-2})\right)
^{n/b^{2}}\left(  b^{-1}\right)  ^{1+2n-2m}\gamma(m-nb^{-2})\prod
_{\substack{k=1-n \\l=1-m \\(k,l)\neq(0,0) }}^{\substack{m-1 \\n-1
}}(lb+kb^{-1})\label{Bmndual}%
\end{equation}
This manifests the duality \cite{Teschner} with respect to $m\leftrightarrow
n$ combined with $b\rightarrow b^{-1}$, $\mu\rightarrow\tilde\mu$ where
\begin{equation}
\pi\tilde\mu\gamma(b^{-2})=\left(  \pi\mu\gamma(b^{2})\right)  ^{1/b^{2}%
}\label{mutilde}%
\end{equation}

\section{Norms of logarithmic primaries}

Take operators $D_{m,n}$ from eq.(\ref{Dmn}) acting in the space of highest
weight vector $\left|  \alpha\right\rangle $ representation of Virasoro
algebra with central charge (\ref{cL}) and dimension $\Delta=\alpha(Q-\alpha
)$. Let the norm be provided by $\left\langle \alpha|\alpha\right\rangle =1$
and $L_{n}=L_{-n}^{\dagger}$. In particular, this defines $D_{mn}^{\dagger}$.
Apparently at $\alpha\rightarrow\alpha_{m,n}$
\begin{equation}
\left\langle \alpha|D_{mn}^{\dagger}D_{m,n}|\alpha\right\rangle =\left(
\alpha-\alpha_{mn}\right)  r_{mn}+o\left(  (\alpha-\alpha_{mn})^{2}\right)
\label{Rmn}%
\end{equation}
where $r_{m,n}$ are coefficients dependent only on $b$. Can we say anything
about $r_{mn}$? Manipulations with Virasoro algebra give quite suggestive
products
\begin{align}
r_{12} &  =-4b^{-1}(1-b^{2})\left(  1+b^{2}\right)  \left(  1+2b^{2}\right)
\nonumber\\
r_{13} &  =24b^{-1}\left(  1-2b^{2}\right)  \left(  1-b^{2}\right)  \left(
1+b^{2}\right)  \left(  1+2b^{2}\right)  \left(  1+3b^{2}\right)
\label{manual}\\
r_{14} &  =-288b^{-1}\left(  1-3b^{2}\right)  \left(  1-2b^{2}\right)  \left(
1-b^{2}\right)  \left(  1+b^{2}\right)  \left(  1+2b^{2}\right)  \left(
1+3b^{2}\right)  \left(  1+4b^{2}\right) \nonumber\\
r_{22} &  =-16b^{-9}\left(  1-b^{2}\right)  ^{2}\left(  1+b^{2}\right)
^{2}\left(  1-2b^{2}\right)  \left(  1+2b^{2}\right)  \left(  2-b^{2}\right)
\left(  2+b^{2}\right) \nonumber\\
&  \ \ldots\nonumber
\end{align}
Is there any general expression?

\section{Poincar\'e disk one-point equation}

Being a local operator identities, eqs.(\ref{heq}) should hold as well for the
Liouville theory in the Poincare\'{ }disk geometry \cite{zz2}, in particular
for the one-point function. The one-point function on a pseudosphere reads
\cite{zz2}
\begin{equation}
U_{p,q}\left(  \alpha\right)  =\frac{\sin\left(  \pi b^{-1}Q\right)
\sin\left(  \pi bQ\right)  \sin\left(  \pi pb^{-1}(Q-2\alpha)\right)
\sin\left(  \pi qb(Q-2\alpha)\right)  }{\sin\left(  \pi b^{-1}pQ\right)
\sin\left(  \pi bqQ\right)  \sin\left(  \pi b^{-1}(Q-2\alpha)\right)
\sin\left(  \pi b(Q-2\alpha)\right)  }U_{1,1}(\alpha)\label{Upq}%
\end{equation}
where
\begin{equation}
U_{1,1}(\alpha)=\frac{\left(  \pi\mu\gamma(b^{2})\right)  ^{-\alpha/b}%
\Gamma(bQ)\Gamma(b^{-1}Q)Q}{\Gamma(b(Q-2\alpha))\Gamma(b^{-1}(Q-2\alpha
))(Q-2\alpha)}\label{U11}%
\end{equation}
We are looking for the following ratio
\begin{equation}
r_{m,n}=2\frac{U_{p,q}(\tilde\alpha_{m,n})}{U_{p,q}(\alpha_{m,n})}%
B_{m,n}\label{rel}%
\end{equation}
It is easy to see that the prefactor in equation (\ref{Upq}) exactly cancels
out in the ratio so that it does not depend on $(p,q)$ and
\begin{equation}
\frac{U_{1,1}(\tilde\alpha_{m,n})}{U_{1,1}(\alpha_{m,n})}=\frac{\left(  \pi
\mu\gamma(b^{2})\right)  ^{-n}}{\left(  mb^{-2}-n\right)  \gamma(m-nb^{2}%
)}\prod_{k=-n}^{n}\left(  mb^{-2}+k\right)  \prod_{l=-m+1}^{m-1}\left(
l+nb^{2}\right) \label{UtildeU}%
\end{equation}
Multiplying this by $2B_{m,n}$ from (\ref{klprod}) we obtain an expression
\begin{equation}
r_{m,n}=2\prod_{\substack{k=1-n \\l=1-m \\(k,l)\neq(0,0)}}^{\substack{m
\\n}}\left(  lb^{-1}+kb\right) \label{Rmnprod}%
\end{equation}
which generalizes the manual results (\ref{manual}). It seems relevant to
compare this product with the denominator in eq.(7.10) of ref.\cite{zz1}.

\section{Application in ``minimal'' gravity}

What such equations can be good for in physical context? Very likely they
might be useful in Liouville gravity (LG) (i.e., the field theory approach to
2D gravity based on the Liouville field theory). To illustrate the idea
consider the following simple example. Let $\left\langle U_{1}\ldots
U_{n}\right\rangle _{\text{LG}}$ be the LG correlation function of any number
of matter primary fields $\Phi_{i}$ ``dressed'' by appropriate Liouville
exponentials $V_{\alpha_{i}}$ so that $U_{i}=\Phi_{i}V_{\alpha_{i}} $. From
(\ref{LFT}) we have
\begin{equation}
\frac\partial{\partial\mu}\left\langle U_{1}\ldots U_{n}\right\rangle
_{\text{LG}}=\int\left\langle U_{1}\ldots U_{n}\;e^{2b\phi}(x)\right\rangle
_{\text{LG}}\;d^{2}x\label{muins}%
\end{equation}
Substituting in the r.h.s $\exp(2b\phi)$ from the basic equation of motion
(\ref{eq11}) one reduces (\ref{muins}) to
\begin{equation}
\frac\partial{\partial\mu}\left\langle U_{1}\ldots U_{n}\right\rangle
_{\text{LG}}=\frac1{\pi\mu b}\int\left\langle U_{1}\ldots U_{n}\;\partial
\bar\partial\phi(x)\right\rangle _{\text{LG}}\;d^{2}x\label{ddphiins}%
\end{equation}
which can be integrated by parts, i.e., turned to contour integrals.
Considering the operator product expansions
\begin{equation}
\phi(x)V_{\alpha}(x^{\prime})=\alpha\log\left|  x-x^{\prime}\right|
V_{\alpha}(x^{\prime})+\ldots\label{phiV}%
\end{equation}
and the asymptotic
\begin{equation}
\phi(x)\sim Q\log\left|  x\right| \label{phiinf}%
\end{equation}
as $\left|  x\right|  \rightarrow\infty$ one summarizes the boundary terms as
\begin{equation}
\frac\partial{\partial\mu}\left\langle U_{1}\ldots U_{n}\right\rangle
_{\text{LG}}=\frac{\left(  \sum_{i=1}^{n}\alpha_{i}-Q\right)  }{\mu
b}\left\langle U_{1}\ldots U_{n}\right\rangle _{\text{LG}}\label{ddmu}%
\end{equation}
The basic Liouville equation of motion allows to carry out explicitly every
(integrated) insertion of the particular operator $U_{I}=\exp(2b\phi)$
reducing it to the correlation function without insertions.

Of course, for more general dressed insertion this trick is not supposed to
work any more. Something exceptional happens in the case of ``minimal''
gravity. The term minimal gravity (MG) stands here for the Liouville gravity
induced by a single minimal CFT model $\mathcal{M}_{p,q}$ ( + ghosts) and then
possibly perturbed by all primary fields of $\mathcal{M}_{p,q}$. Unlike
general Liouville gravity, MG is expected to be exactly solvable. This is
because the matrix model approach to 2D gravity (for a review see e.g.,
\cite{M}) provides explicit expressions for many observables in what is
believed to be equivalent to MG.

In Liouville gravity it's reasonable to start with unperturbed (conformal)
matter. Conformal matter doesn't interact with the background except for
through the conformal anomaly. The Liouville gravity is decoupled to pure
matter CFT, Liouville and ghosts. In minimal $p,q$ gravity the matter central
charge is
\begin{equation}
c_{\text{M}}=1-6(b^{-1}-b)^{2}\label{cM}%
\end{equation}
where $b=\sqrt{p/q}$. We keep the same notation $b$ for this parameter as for
the Liouville parameter in the previous sections because it's in fact the one.
For, the Liouville central charge (\ref{cL}) correctly adds up with
(\ref{cM})
\begin{equation}
c_{\text{M}}+c_{\text{L}}=26\label{cadd}%
\end{equation}
In principle $b^{2}$ is a rational number but for what follows it doesn't
really matter, it can arbitrary.

Possible perturbations are the fields $\Phi_{m,n}$ from the spectrum of
$\mathcal{M}_{p,q}$. They have dimensions \cite{BPZ}
\begin{equation}
\Delta_{m,n}^{\text{(M)}}=\frac{\left(  b^{-1}m-bn\right)  ^{2}-\left(
b^{-1}-b\right)  ^{2}}4=1-\tilde\Delta_{m,n}\label{DeltamnM}%
\end{equation}
with $\tilde\Delta_{m,n}=\tilde\alpha_{m,n}\left(  Q-\tilde\alpha
_{m,n}\right)  $ and $\tilde\alpha_{m,n}$ from eq.(\ref{alphatilde}). Thus the
Liouville field $\tilde V_{m,n}$ of sect.4 can be used as the ``gravitational
dressing'' of $\Phi_{m,n}$ and the perturbing operator of dimension $(1,1)$
is
\begin{equation}
U_{m,n}(x)=\Phi_{m,n}\tilde V_{m,n}(x)\label{Umn}%
\end{equation}
It is one of the peculiarities of MG: all matter fields are ``dressed'' by
Liouville exponentials entering the higher equations of motion (\ref{heq}).

Our first goal is to learn to integrate such insertions
\begin{equation}
\int\left\langle U_{m,n}(x)\;\ldots\right\rangle _{\text{MG}}d^{2}%
x\label{Umnint}%
\end{equation}
where $\left\langle \ldots\right\rangle _{\text{MG}}$ stands for the joint
correlation function of matter, Liouville and ghosts, and $\ldots$ is for any
observable. Higher equations of motion (\ref{heq}) allow to substitute the
Liouville part of this insertion as
\begin{equation}
\int\left\langle U_{m,n}(x)\;\ldots\right\rangle _{\text{MG}}d^{2}%
x=\frac1{B_{m,n}}\int\left\langle \Phi_{m,n}\bar D_{m,n}D_{m,n}V_{m,n}%
^{\prime}(x)\;\ldots\right\rangle _{\text{MG}}d^{2}x\label{heqsub}%
\end{equation}
What is also specific for MG is that the matter fields $\Phi_{m,n}$ are all
degenerate \cite{BPZ}
\begin{equation}
D_{m,n}^{\text{(M)}}\Phi_{m,n}=\bar D_{m,n}^{\text{(M)}}\Phi_{m,n}%
=0\label{BPZnull}%
\end{equation}
where $D_{m,n}^{\text{(M)}}$ is the matter version of the Liouville operator
$D_{m,n}$. In fact, $D_{m,n}^{\text{(M)}}$ is explicitly obtained from
$D_{m,n}$ by replacing all $L_{n}$ with the matter conformal generators
$L_{n}^{\text{(M)}}$ and substituting $b^{2}\rightarrow-b^{2}$. This allows to
define the joint operators
\begin{equation}
\mathcal{D}_{m,n}=D_{m,n}-D_{m,n}^{\text{(M)}}\label{Dmnjoint}%
\end{equation}
and write
\begin{equation}
\int\left\langle U_{m,n}(x)\;\ldots\right\rangle _{\text{MG}}d^{2}%
x=\frac1{B_{m,n}}\int\left\langle \mathcal{\bar D}_{m,n}\mathcal{D}%
_{m,n}\Theta_{m,n}^{\prime}(x)\;\ldots\right\rangle _{\text{MG}}%
d^{2}x\label{Thetaprim}%
\end{equation}
where $\Theta_{m,n}^{\prime}=\Phi_{m,n}V_{m,n}^{\prime}$.

A very likely statement is that if $\ldots$ in this correlation function is
BRST closed, the integrand in the right hand side of (\ref{Thetaprim}) is a
derivative in $x$ and $\bar x$%
\[
\mathcal{\bar D}_{m,n}\mathcal{D}_{m,n}\left(  \Theta_{m,n}^{\prime}\right)
=\partial\bar\partial\left(  \bar H_{m,n}H_{m,n}\Theta_{m,n}^{\prime}\right)
+\text{BRST exact}
\]
and thus can be reduced to boundary terms. Here $H_{m,n}$ are operators of
level $mn$ and ghost number $0$ constructed from $L_{n}$, $L_{n}^{\text{(M)}}$
and ghosts. Few such operators were explicitly derived in \ refs.\cite{W, Bow,
Imbimbo}.

Let's present here an explicit calculation for the simplest example
$(m,n)=(1,2)$. In this case
\begin{equation}
\mathcal{D}_{1,2}=\partial_{\text{L}}^{2}-\partial_{\text{M}}^{2}%
+b^{2}\mathcal{L}_{-2}\label{D12joint}%
\end{equation}
where $\mathcal{L}_{n}=L_{n}+L_{n}^{\text{(M)}}$. It is verified
straightforwardly that
\begin{align}
\  &  \mathcal{\bar{D}}_{1,2}\mathcal{D}_{1,2}O_{12}^{\prime}-\partial
\bar{\partial}\left(  \bar{H}_{12}H_{12}\Theta_{12}^{\prime}\right)
=\label{QQcomm}\\
&  \ b^{4}\{\mathcal{\bar{Q},}\{\mathcal{Q},B\bar{B}\Theta_{12}^{\prime
}\}\}-b^{2}\left\{  \mathcal{Q},\bar{\partial}\left(  \bar{H}_{12}B\Theta
_{12}^{\prime}\right)  \right\}  -b^{2}\left\{  \mathcal{\bar{Q}}%
,\partial\left(  H_{12}\bar{B}\Theta_{12}^{\prime}\right)  \right\} \nonumber
\end{align}
Here $B$ and $C$ are ghost fields (we use here the unusual upper case letters
for the usual ghosts for not to mix $B$ with the parameter $b$ of minimal
gravity), $\mathcal{Q}$ is the BRST charge
\begin{equation}
\mathcal{Q}=\oint\left(  C(T_{\text{L}}+T_{\text{M}})+C\partial CB\right)
\frac{dz}{2\pi i}\label{brst}%
\end{equation}
and operator $H_{12}$ reads explicitly \cite{Bow, Imbimbo}
\begin{equation}
H_{12}=\partial_{\text{M}}-\partial_{\text{L}}+b^{2}CB\label{H12}%
\end{equation}
Notice, that in the usual ground ring constructions \cite{W, KMS} (see also
\cite{All} for more recent developments) $H_{m,n}$ is applied to $\Theta
_{m,n}=\Phi_{m,n}V_{m,n}$ to obtain the ground ring element
\begin{equation}
O_{m,n}=H_{m,n}\bar{H}_{m,n}\Theta_{m,n}\label{Omn}%
\end{equation}
Now it is applied to the product $\Phi_{12}V_{12}^{\prime}$ involving the
logarithmic degenerate field of Liouville theory.

If we introduce also the operator $\hat{Q}$ acting on fields as
\begin{equation}
\hat{Q}f=\left\{  \mathcal{Q},f\right\} \label{Qhat}%
\end{equation}
eq.(\ref{QQcomm}) can be rewritten in quite a compact and suggestive form
\begin{equation}
\mathcal{\bar{D}}_{1,2}\mathcal{D}_{1,2}O_{12}^{\prime}=\left(  \hat{Q}%
B-b^{2}\partial H_{12}\right)  \left(  \hat{\bar{Q}}\bar{B}-b^{2}\bar
{\partial}\bar{H}_{12}\right)  O_{12}^{\prime}\label{suggestive}%
\end{equation}
where $\partial$ and $\bar{\partial}$ inside the brackets act on everything to
the right of them.

\textbf{Acknowledgments}

I'm grateful to A.Zamolodchikov, A.Neveu and D.Kutasov for encouraging
interest to the work and many useful comments. The work was supported by the
European Commitee under contract HPRN-CT-2002-00325.

\end{document}